# REVISITING BLUESHIFT INTERPRETATION IN LIGHT OF RECENT DISCOVERY OF MULTIPLE SYSTEMS OF QUASARS

*J. Singh and S. Haque

Department of Physics, University of the West Indies, St. Augustine, Trinidad

*Corresponding Author - Email: justin.singh1@my.uwi.edu, Address: #20 Amrit Street, D'Abadie, Trinidad and Tobago, West Indies

**Abstract**

This study investigates the anomalies associated with redshifts from emission lines in certain quasar candidates and the viability of a blueshift interpretation instead. The sample was taken from the Million Quasars Catalog (MILLIQUAS), representing the unidentified class with a redshift greater than 1. This sample was further constrained to those with spectra available, giving 208 candidates in total. This paper presents results on the sample, with the reported redshifts and the proposed blueshift interpretation. A subset of 38% of the sample was further analyzed using the best redshift interpretation of the emission lines from our analysis, which differed from the reported redshifts, in comparison with the blueshift interpretation. The number of unidentified lines under each interpretation was compared and was found to be statistically different with a P-value < 0.05, with a larger number of unidentified lines under the redshift interpretation. The average spread values were also compared and found to be statistically different with a P-value < 0.05, with blueshift having the smallest spread. Eighty-eight percent (88%) of the analyzed sample, that is 183 quasar candidates, provided an overall better interpretation under the blueshift hypothesis, 9%, which is 19 candidates, had a better interpretation under the redshift hypothesis and 3%, which is 6 candidates, had no identifiable lines. This indicates the importance to consider this possibility as well in light of new discoveries such as the discovery of multiple quasar systems that can lead to ejections, which has implications for the dynamics of quasars and the line of sight.

## 1) Introduction

Since the discovery of quasars in the 1960s (Greenstein 1963, Matthews & Sandage 1962) there has been contention about the nature of their redshift especially in the early days (Haque, 2010). The cause of redshifts can be gravitational or cosmological being relativistic in nature due to the high velocities implied. It is now commonly accepted that gravitational redshift is negligible in the redshift of quasars (Narlikar & Edmunds 1981). It is mostly attributed to cosmological in nature with some relativistic effects being independent of the cosmological. The cosmological effect is due to the expansion of the universe and the other relativistic effect, due to some motion from the observer that is independent of the expansion of the universe. Motion away from the observer shows redshifting of the emission lines while motion towards the observer will show blueshifting of the lines (Haque-Copilah et al 1997). Most cosmological objects' spectra are analysed using a redshift interpretation, and quasars are no different. However, it is possible for scenarios where there is ejection of a quasar involved in a multi-body system and depending on the line of sight, this will generate blueshift in the emission lines (Basu & Haque-Copilah, 2001; Basu et al 2000).

Quasars sometimes show blueshifts in certain emission lines or their Broad Line Regions (BLR), which is attributed to their rotating nature or the direction of their relativistic jet (Haque & Valtonen 2008), but blueshift of the quasar due to its motion itself is not considered. The high ionization Broad Emission Lines (BELs) are believed to be produced closer to the central black hole and they are found to be significantly blueshifted compared to the low ionization BELs (Sun, 2018). It is widely believed these blueshifts are a consequence of orientation, however Mouyuan Sun (2018) study showed that this theory is not favoured. Some quasars show a blueshift of the BLR with respect to the systematic redshift and Narrow Line Region (NLR) and it is believed that this occurs due to the quasar being a recoiling black hole from the merger of two Supermassive black holes (SMBHs) (Marco Chiaberge 2018).

Line identification of quasar spectra is programmed to match observed lines with lines on the red side of the spectrum, and therefore can only determine possible redshifts without considering blueshifts (Basu 2004). Evidence shows that for some cosmic sources the spread, where spread is the difference between the maximum and minimum line values, of redshift values exceeds the acceptable spread of 0.01 for redshift, whereas the possible corresponding blueshift values have less of a spread (Basu 2000). There is also an acceptance of redshifts which give lines that do not correspond to their expected strengths, for example, the BELs Lyα, CIV, CIII] and MgII being identified for weak and/or narrow lines and otherwise weak lines being identified as strong observed lines. All observed lines are required to also have the same redshift value (Basu 2010). This contradicts the observation of the BLR being blueshifted, both with regards to the systematic redshift and the NLR. This lack of consideration of blueshifting of emission lines can account for some of the inaccuracies seen with redshift interpretation is considered.

Blueshifting of lines is a plausible alternative to the redshift interpretations of some spectra based on the dynamics of the system. The ejection mechanism in multi-body systems can give rise to the blueshift of the BLR. This scenario can be applied to both interpretations. Given the three dimensional nature of space, it is equally probable that the ejected bodies will be towards the observer at an angle such that the blueshift of the object is larger than the redshift due to the expansion of the universe (Popowski & Weinzierl 2004). It should be noted that due to the ratio of velocities in the calculation of doppler shift, blueshift can only have a maximum value of -1, as any magnitude greater will entail a velocity greater than the speed of light. This blueshift value does not factor in the cosmological redshift and therefore any overall blueshift value would need to be smaller in magnitude, i.e. simply that some part of the spectra may be blueshifted as well, if their velocity toward the observer is greater than the expansion of the universe. Achieving such a velocity is not impossible as multi-body problems have shown outcomes where one of the bodies was ejected at high velocities from the system. Chitan (2022) shows that 25% of three-body interactions between Supermassive Black Holes (SMBH) end in high-speed ejections. These ejections will demonstrate blueshifts if they are directed towards the observer.

Van Dokkum et al (2023) has recently observed what is likely an ejected SMBH from a triple system. This ejected SMBH was found to be moving at a very high velocity, and while not at a direct angle towards us, the observer, or near the maximum ejection velocity of 0.94c (Narlikar & Edmunds 1981), this is prominent evidence of ejected SMBHs as expected in 25% of triple systems (Chitan et al 2022).

Farina et al (2013) reported on a triple quasar system and demonstrated that despite a low probability of detection, triple quasar systems are still discovered. Three-body systems among black holes, specifically SMBHs are even more commonly observed now, with 15 systems currently confirmed (Foord et al 2021). More are likely to be discovered due to the detection of gravitational waves now, which are generated in such interactions.

Given these recent works, we feel that it is time to revisit the blueshift interpretation of quasars' spectra as an alternative to redshift interpretation that are not well defined.

2) **Methodology**

The data was obtained from the Million Quasars Catalogue (MILLIQUAS), which consists of information from a wide range of other databases, such as VIPERS, and SDSS. The unidentified class was chosen with the constraint of z>1, since z=1 is approximately equal to half the age of the universe, z>1 would provide quasar candidates in the earlier universe where we expect more multiple body interactions as the universe is younger and the objects closer in proximity. This yielded 645 possible quasars, they were then subjected to ESO Archive Science Portal to obtain spectra. Of the 645 quasars, 208 had available spectra, all from the VIPERS catalogue within the wavelength range 551.4 nm to 948.4 nm. This list of 645 candidates was also subjected to the SDSS Science Archive Server where some spectra was obtained, however they also had spectra obtained from the ESO portal. Given the small (approximately 10) amount of spectra obtained from the SDSS portal, the ESO spectra was used instead for consistency among the other spectra.

Using a combination of Python packages (specutils, astropy, numpy and matplotlib), the spectra were analysed to identify the emission lines present in the spectra by removing the noise present using a generated spectra. The emission lines were then subjected to a MATLAB function, specifically written to identify the lines under the redshift hypothesis by matching the observed wavelengths with rest wavelengths of emission lines. The code accepts the redshift as an input, as well as the observed lines, from there it produces the identified lines within an acceptable range and gives the average redshift value as well as the redshift spread.

Thirty-eight percent of the spectra were analysed to determine the best redshift interpretation independently, regardless of the reported redshift to ensure the most accurate interpretation possible. The first emission line for these candidates were identified initially by cycling through the list of possible emission line and then the remaining emission lines were identified based on this. The different interpretations were then manually analyzed to determine which gave the best interpretation of the line strengths as well as identifies the most emission lines with the smallest spread.

The remainder were analysed using the reported redshift and constrained to the acceptable redshift spread of 0.02. The emission lines were then analysed using a second MATLAB function, specifically designed to identify the lines under the blueshift hypothesis to give the best blueshift interpretation. This function worked similarly to the redshift function that first determined the first emission line, however, instead of cycling through all the possible emission lines, the first emission was manually assigned based on the line strength and the function took this input and assigned the other emission lines. From all the different interpretations, the one with the best aligned line strengths within the acceptable spread was accepted.

The spectral lines used as search lines were a combination of two lists: a list compiled by Chojnowski (2010) using data from the National Institute of Standards and Technology and the list from Basu (2010) shown in Table 1. The list by Chojnowski (2010) are lines commonly found in galaxies, AGNs and Quasars, however, there are no relative line strengths presented, thus line strengths were calculated by utilizing line strength ratios of quasars with respect to emission lines with known strengths. Some of the lines presented by Basu (2010) do not correspond to exact wavelengths of lines found in the National Institute of Standards and Technology (NIST), however, both wavelengths are relatively close to each other, within a few angstroms, making the difference too small to be significant.

**Table 1. Composed list of the search lines used to identify the emission lines observed in quasar spectra**.

| Redshift search lines | | Blueshift search lines | |
|---|---|---|---|
| Emission Line(Å) | Relative Strength | Emission Line(Å) | Relative Strength |
| MgX 615 | 0.1 | HeI 5876 | 0.5 |
| ArV763 | 0.1 | [FeVII] 6087 | 0.1 |
| NIV 765 | 0.1 | [OI] 6300 | 1.0 |
| NeVII 774 | 1.0 | [SIII] 6312 | 0.2 |
| OIII 834 | 0.1 | SiIII 6335 | 0.1 |
| Ly$\eta$ 926 | 0.1 | OI 6364 | 0.5 |
| Ly$\zeta$ 931 | 0.1 | FeX 6374 | 0.1 |
| Ly$\epsilon$ 938 | 0.1 | [NV] 6548 | 0.5 |
| SVI 944 | 0.2 | H$\alpha$ 6563 | 2.0 |
| Ly$\delta$ 950 | 0.2 | [NII] 6584 | 1.5 |
| Ly$\gamma$ 973 | 0.2 | HeI 6678 | 0.1 |
| CIII 977 | 0.2 | [SII] 6716 | 0.5 |
| NIII] 991 | 0.1 | [SII] 6730 | 0.5 |
| Ly$\beta$ 1026 | 0.5 | [ArV] 7006 | 0.2 |
| OVI 1034 | 0.5 | HeI 7065 | 0.5 |
| SiVI 1072 | 0.1 | [ArIII] 7136 | 0.2 |
| NII 1085 | 0.1 | [FeII] 7155 | 0.1 |
| Ly$\alpha$ 1216 | 2.0 | [OII] 7324 | 0.5 |
| OV] 1218 | 0.1 | [NiII] 7378 | 0.1 |
| NV 1240 | 0.1 | [NiII] 7411 | 0.1 |
| SII 1263 | 0.1 | [SXII] 7611 | 0.5 |
| OI 1303 | 0.2 | [ArIII] 7751 | 1.0 |
| CII 1335 | 0.2 | OI 7773 | 0.2 |
| SiIV 1400 | 0.1 | FeXI 7892 | 0.1 |
| OIV] 1406 | 0.1 | HeII 8237 | 0.1 |
| NIV 1486 | 0.1 | OI 8449 | 1.0 |
| CIV] 1549 | 2.0 | CII 8489 | 0.1 |
| NeV] 1575 | 0.1 | CaII 8542 | 0.1 |
| NeIV] 1602 | 0.1 | [FeII] 8617 | 0.1 |
| HeII 1640 | 0.5 | CaII 8662 | 0.1 |
| OIII] 1664 | 0.2 | NI 8712 | 0.1 |
| NiIII 1719 | 0.1 | [SIII] 9069 | 0.5 |
| NiII 1742 | 0.1 | P$\xi$ 9240 | 0.1 |
| NIII] 1750 | 0.1 | [SIII] 9532 | 1.0 |
| SiII 1814 | 0.1 | P$\epsilon$ 9552 | 0.1 |
| AlIII 1857 | 1.0 | [CI] 9824 | 0.1 |
| [SiIII] 1892 | 0.1 | [CI] 9850 | 1.0 |
| [CIII] 1909 | 2.0 | SVIII 9912 | 0.1 |
| HeI 2058 | 0.1 | HeI 10049 | 0.2 |
| NII] 2143 | 0.1 | SII 10124 | 1.0 |
| [OIII] 2321 | 0.1 | [SII] 10287 | 0.1 |
| CII 2326 | 0.2 | [NI] 10300 | 0.1 |
| OII 2476 | 0.1 | HeI 10400 | 0.1 |
| [MgVII] 2506 | 0.1 | [FeXIII] 10747 | 0.2 |
| [MgVII] 2526 | 0.1 | HeI 10830 | 1.0 |
| [FeXI] 2649 | 0.1 | MgII 10926 | 0.1 |
| MgII 2798 | 2.0 | P$\gamma$ 10938 | 0.5 |
| MgI 2852 | 0.1 | OIV 11210 | 0.1 |
| [ArIV] 2854 | 0.5 | OIV 12287 | 0.1 |
| [ArIV] 2869 | 1.0 | SIX 12523 | 0.1 |
| MgV 2931 | 0.2 | FeII 12567 | 0.1 |
| [NeV] 2973 | 0.5 | FeII 12786 | 0.1 |
| OIII 3133 | 0.2 | P$\beta$ 12818 | 1.0 |
| HeI 3188 | 0.1 | FeII 12970 | 0.1 |
| HeII 3204 | 0.2 | FeII 13003 | 0.1 |
| OIII 3312 | 0.1 | [FeII] 15995 | 0.1 |
| [NeV] 3426 | 1.0 | [FeII] 16440 | 0.1 |
| OIII 3444 | 0.1 | [FeI] 16770 | 0.1 |
| [FeVII] 3586 | 0.1 | HeI 17008 | 0.2 |
| [FeVI] 3663 | 0.1 | P$\alpha$ 18751 | 2.0 |
| [OII] 3727 | 1.0 | Br$\delta$ 19445 | 0.1 |
| [NeIII] 3869 | 1.0 | H2 20338 | 0.1 |
| H$\xi$ 3889 | 0.1 | HeI 20580 | 0.1 |
| H$\epsilon$ 3968 | 0.1 | H2 20587 | 0.1 |
| SII 4069 | 0.1 | H2 20735 | 0.1 |
| H$\delta$ 4102 | 0.2 | H2 21218 | 0.2 |
| [FeII] 4287 | 0.1 | H2 21542 | 0.1 |
| H$\gamma$ 4340 | 1.5 | Br$\gamma$ 21660 | 0.1 |
| [OIII] 4363 | 1.5 | H2 22233 | 0.1 |
| HeI 4471 | 0.1 | H2 22477 | 0.1 |
| MgII 4481 | 0.1 | [CaVIII] 23210 | 1.0 |
| CIII 4649 | 0.1 | [SiVII] 24830 | 1.0 |
| H$\beta$ 4861 | 1.5 | Br$\beta$ 26250 | 0.5 |
| HeI 4921 | 0.1 | [SIX] 29346 | 0.2 |
| [OIII] 4959 | 1.0 | Br$\alpha$ 40500 | 0.5 |
| [OIII] 5007 | 2.0 | | |
| [NI] 5199 | 0.1 | | |
| [FeVII] 5276 | 0.1 | | |
| [CaV] 5301 | 1.0 | | |
| SII 5454 | 0.1 | | |

Hypothesis testing with a paired sample z-test was used to compare the number of unidentified lines for each interpretation and the mean spread for both redshift and blueshift to test the possibility of a significant difference between the two interpretations. The equation utilized across the MATLAB functions is given by

$$z = \frac{\lambda_o - \lambda_e}{\lambda_e}$$

Where z is the shift value, $\lambda_o$ is the observed wavelength of the emission line and $\lambda_e$ is the emitted or rest wavelength of the emission line.

3) **Results**

From the analysis, 9% of those analysed have a better interpretation under the redshift hypothesis and 3% had no identifiable emission lines. The remaining 88% demonstrated better interpretations under the blueshift hypothesis. Search line strengths aligning with their observed strengths and a lower spread both constitute a better interpretation, however the number of observed lines that could not be identified also factored into the suitability of the interpretation. From the 88% that had a better blueshift interpretation, three quasar candidates, VIPERS 113187213, VIPERS 114149180 and VIPERS 410050999, had more unidentified lines under the blueshift hypothesis than the redshift.

However, due to the difference in the spread being significantly different, blueshift spread being 0.0243, 0.0196 and 0.0277 less than redshift spread respectively, and the identified line strengths fitting the emission line strengths better for the blueshift than redshift, the better interpretation was deemed to be the blueshift hypothesis. Twenty-one other quasar candidates, listed in Table 2, had smaller spreads for redshift and similar or more unidentified lines, however, these candidates gave strengths more in conjunction with the identified line strengths with the blueshift hypothesis and given the difference between the spreads were 0.0065 or less, these candidates were deemed to have a better interpretation under the blueshift hypothesis.

**Table 2. Special cases of Quasar Candidates.**

| Name | RA | Dec | Redshift |
|---|---|---|---|
| VIPERS 410093351 | 22:12:42.00 | 02:10:43.2 | 2.052 |
| VIPERS 410110039 | 22:13:58.07 | 02:15:24.7 | 2.919 |
| VIPERS 403194183 | 22:08:33.93 | 01:42:05.4 | 1.491 |
| VIPERS 403114254 | 22:11:15.06 | 01:20:28.9 | 2.172 |
| VIPERS 403063680 | 22:09:11.49 | 01:06:58.6 | 3.455 |
| VIPERS 403015810 | 22:11:09.69 | 00:53:26.2 | 1.159 |
| VIPERS 402140488 | 22:04:59.74 | 01:23:32.7 | 1.152 |
| VIPERS 402089423 | 22:06:38.01 | 01:11:22.8 | 1.465 |
| VIPERS 402082334 | 22:07:03.18 | 01:09:35.6 | 1.386 |
| VIPERS 402054002 | 22:07:36.54 | 01:02:13.3 | 1.743 |
| VIPERS 401028814 | 22:01:58.15 | 00:56:32.6 | 1.239 |
| VIPERS 126087074 | 02:30:11.87 | -04:13:56.5 | 1.638 |
| VIPERS 126052925 | 02:28:26.87 | -04:24:13.6 | 2.078 |
| VIPERS 126022365 | 02:28:17.56 | -04:34:10.3 | 3.305 |
| VIPERS 411134365 | 22:18:20.28 | 02:16:05.2 | 1.325 |
| VIPERS 404014685 | 22:11:41.57 | 00:53:10.8 | 1.428 |
| VIPERS 403198919 | 22:07:59.10 | 01:43:26.7 | 1.433 |
| VIPERS 403198022 | 22:10:38.63 | 01:43:08.2 | 1.298 |
| VIPERS 403146598 | 22:08:45.17 | 01:29:19.8 | 1.192 |
| VIPERS 403139450 | 22:08:21.18 | 01:27:20.6 | 1.999 |
| VIPERS 403085027 | 22:08:51.80 | 01:12:34.9 | 1.925 |

Two candidates (VIPERS 403114254 and VIPERS 403194183) possessed blueshift spreads 0.0069 greater than their redshift spreads, however, the number of unidentified lines were much greater under the redshift hypothesis and that, in conjunction with the better alignment of line strengths under the blueshift hypothesis, led to them being considered as better blueshift interpretations.

VIPERS 403114254 had six unidentified lines under the redshift hypothesis, with the two identified lines being the CII] 1909 and MgII 2798 lines, one of which was shown as a weak line when both are known as strong lines. Under the blueshift hypothesis, it indicated one unidentified line with minute changes in line strength, such as, [SII] 6716 being a traditionally mid strength being shown as weak and SVI 9912 being a traditionally weak line being identified as a mid-strength line. VIPERS 403194183 had 3 unidentified lines under the redshift hypothesis, with all the 3 identified lines aligning with their respective line strengths. There were no unidentified lines under the blueshift hypothesis with all the line strengths aligning with their observed strengths.

Of the quasars that were classified to have a better interpretation under blueshift, only 18% had blueshift spreads greater than the acceptable spread of 0.02. Table 3 goes further to show the mean spread for the redshift and blueshift interpretation for the entire sample. The redshift spread is shown as 0.0295 and the blueshift spread is given by 0.0141, indicating the mean redshift spread to be larger than the acceptable spread of 0.02 and the mean blueshift spread to be less than it. Table 4 indicates the mean spreads are statistically different at a 0.05 level of significance.

**Table 3. Comparisons between the two paired samples.**

|  | Mean | Std. Deviation |
|---|---|---|
| Unidentified lines for redshift | 2.5248 | 2.0129 |
| Unidentified lines for blueshift | 0.7030 | 0.8758 |
| Redshift spread | 0.0295 | 0.0294 |
| Blueshift spread | 0.0141 | 0.0094 |

The number of unidentified lines were also investigated. Table 3 shows the mean number of unidentified lines for redshift and blueshift which were 2.52 and 0.72 respectively. As indicated by Table 4, the mean number of unidentified lines were statistically different at 0.05 level of significance for redshift and blueshift. Table 3 also shows the large difference between the standard deviation for redshift number of unidentified lines and blueshift number of identified lines, indicating the inconsistent nature of redshift number of unidentified lines showing it can be as much as four unidentified lines under the redshift interpretation.

**Table 4. Two paired samples are significantly different.**

| Difference between: | z value | 2-tailed Significance |
|---|---|---|
| Redshift and blueshift Unidentified lines | 11.695 | 0.000 |
| Redshift and blueshift spread | 7.080 | 1.44E-12 |

4) **Discussion**

In this paper we investigated the discrepancies in the redshift interpretation of quasar candidates by conducting redshift interpretation tests to compare to reported redshifts for 38% of quasar candidates of the unidentified class of the MILLIQUAS catalogue. A blueshift interpretation test was also conducted on all of the candidates to determine if the emission line inconsistencies seen with the redshift interpretation can be better explained with the blueshifting of emission lines. The best interpretation was determined using the factors of aligned line strength, number of unidentified lines and the spread of the redshift or blueshift.

Blueshift interpretation was found to be a much better fit for most of the sample. This highlights that there are inconsistences among some redshift interpretations, particularly the ones of quasars whose classification are unclear and perhaps it is just for this reason.

Having high S/N is important to ensure that all observed lines are indeed emission lines and to ensure emission lines are not lost in the noise as it was noted that 3% of the sample could not be analysed as there were no emission lines derived from the spectra once the noise spectrum was accounted for.

Tables 5, 6 and 7 show examples of quasar candidates that displayed a better interpretation under the blueshift interpretation rather than the redshift interpretation. Table 5 and Figure 1 displays the observed lines and the emitted lines under both redshift and blueshift hypothesis for VIPERS 403194183.

**Table 5 showing emission lines for VIPERS 403194183**

| VIPERS 403194183 | | | |
|---|---|---|---|
| | Observed lines (Å) | Redshift (Å) | Blueshift (Å) |
| A | 6892 | --- | P$\xi$ |
| B | 6957 | Mg II | [S III] |
| C | 7085 | Mg I | P$\epsilon$ |
| D | 7721 | --- | [N I] |
| E | 7749 | O III | He I |
| F | 7863 | --- | [Fe XIII] |

**Figure 1 showing emission lines for VIPERS 403194183**

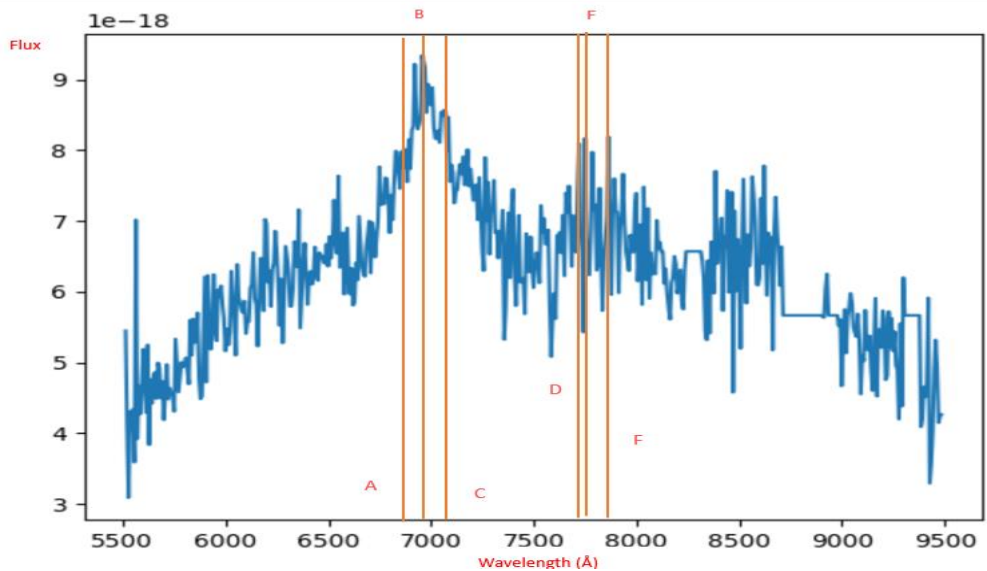

VIPERS 403194183 had a total of six observed lines with one strong line and five weak or semi-weak lines. Under the redshift hypothesis, the strong line was identified as MgII 2798 which is also a strong line given by B in Figure 1. One weak and one semi-weak line was also identified as lines with appropriate strengths, however, the other three lines were unidentified under this interpretation. The blueshift hypothesis identified all observed lines with emitted lines of similar strengths, which indicated the blueshift hypothesis gave a better explanation of the observed lines.

Table 6 and Figure 2 displays the interpretations of the observed lines for VIPERS 403198919. VIPERS 403198919 had four observed lines, with one strong and three weak lines. Under the redshift interpretation the strong line was identified as Mg II 2798 and one of the weak lines as $H\xi$. Another weak line was identified as [Ne III] 3869 which is a strong line, given by B in Figure 2, and the last weak line was left unidentified. The blueshift interpretation identified all the observed lines with emitted lines of the appropriate strengths, the strong line as [S III] 9532 and the weak lines as S IX 12523, Fe II 12567 and Fe II 12786.

**Table 6 showing emission lines for VIPERS 403198919**

| VIPERS 403198919 | | | |
|---|---|---|---|
| | Observed lines (Å) | Redshift (Å) | Blueshift (Å) |
| A | 6809 | Mg II | [S III] |
| B | 9306 | [Ne III] | S IX |
| C | 9327 | --- | Fe II |
| D | 9434 | Hξ | Fe II |

**Figure 2 showing emission lines for VIPERS 403198919**

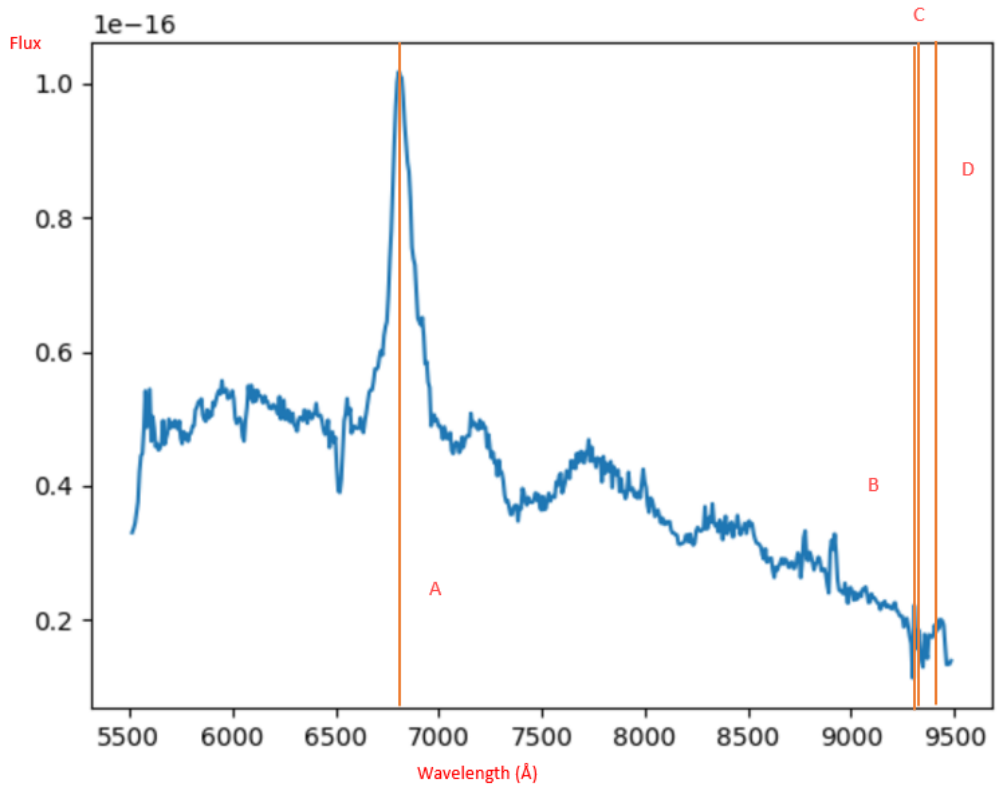

Table 7 and Figure 3 displays VIPERS 403093966 and its observed and emitted emission lines. VIPERS 403093966 had eight observed emission lines, one strong line and seven weak lines. The redshift interpretation appropriately identified the strong line and one weak line, but the remaining weak lines were left unidentified. The blueshift interpretation left only one line unidentified, appropriately identifying the strong line and six weak lines with lines of same strengths.

**Table 7 showing emission lines for VIPERS 403093966**

| VIPERS 403093966 | | | |
|---|---|---|---|
| | Observed lines (Å) | Redshift (Å) | Blueshift (Å) |
| A | 5857 | Mg II | O I |
| B | 7278 | --- | [S II] |
| C | 7314 | --- | [N I] |
| D | 7328 | --- | He I |
| E | 7349 | --- | --- |
| F | 7599 | [Fe VI] | [Fe XIII] |
| G | 7913 | --- | O I |
| H | 8827 | --- | S IX |

**Figure 3 showing emission lines for VIPERS 403093966**

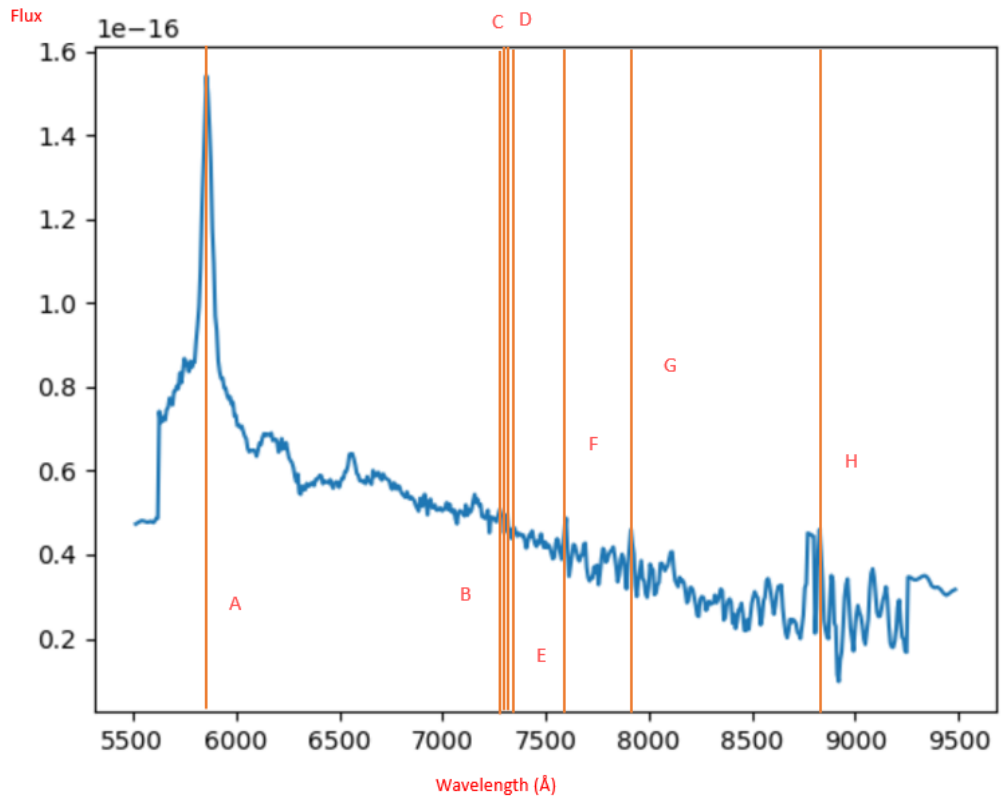

Tables 8 and 9 displays VIPERS 404032519 and VIPERS 403054867 respectively. VIPERS 404032519, whose spectra is seen in Figure 4, had four observed emission lines, one strong and three weak lines. Both redshift and blueshift interpretations identified all observed lines, however, the blueshift emitted lines had a spread of 0.0321 and the redshift had a spread of 0.0216. This indicated a better interpretation under the redshift hypothesis. A point of note is that both spreads are larger than the acceptable spread of 0.01.

**Table 8 showing emission lines for VIPERS 404032519**

| VIPERS 404032519 | | | |
|---|---|---|---|
| | Observed lines | Redshift (Å) | Blueshift |

|   | (Å) |   | (Å) |
|---|---|---|---|
| A | 6421 | Mg II | He I |
| B | 8970 | Hξ | [Fe II] |
| C | 9320 | S II | [Fe II] |
| D | 9484 | Hδ | [Fe I] |

**Figure 4 showing emission lines for VIPERS 404032519**

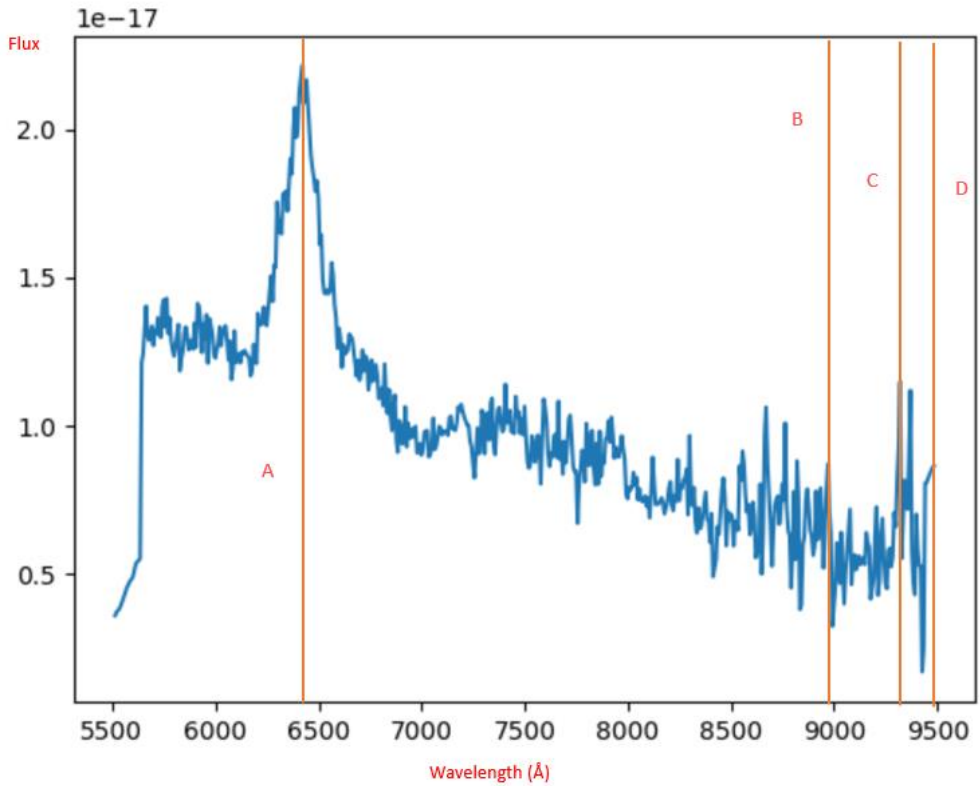

VIPERS 403054867 had only two observed emission lines as seen in the spectra in Figure 5, one strong and one weak. Both were identified with emitted lines of the appropriate strength under both the redshift and blueshift hypotheses, Mg II 2798 and [S III] 9532 for the strong line and O III 3312 and O I 11210 for the weak line respectively. Both also displayed a spread within the acceptable range of 0.01, however, the redshift interpretation had a smaller spread and so it was classified as a candidate with a better interpretation under the redshift hypothesis.

**Table 9 showing emission lines for VIPERS 403054867**

| VIPERS 403054867 | | | |
|---|---|---|---|
| | Observed lines (Å) | Redshift (Å) | Blueshift (Å) |
| A | 8020 | Mg II | [S III] |
| B | 9484 | O III | O I |

**Figure 5 showing emission lines for VIPERS 403054867**

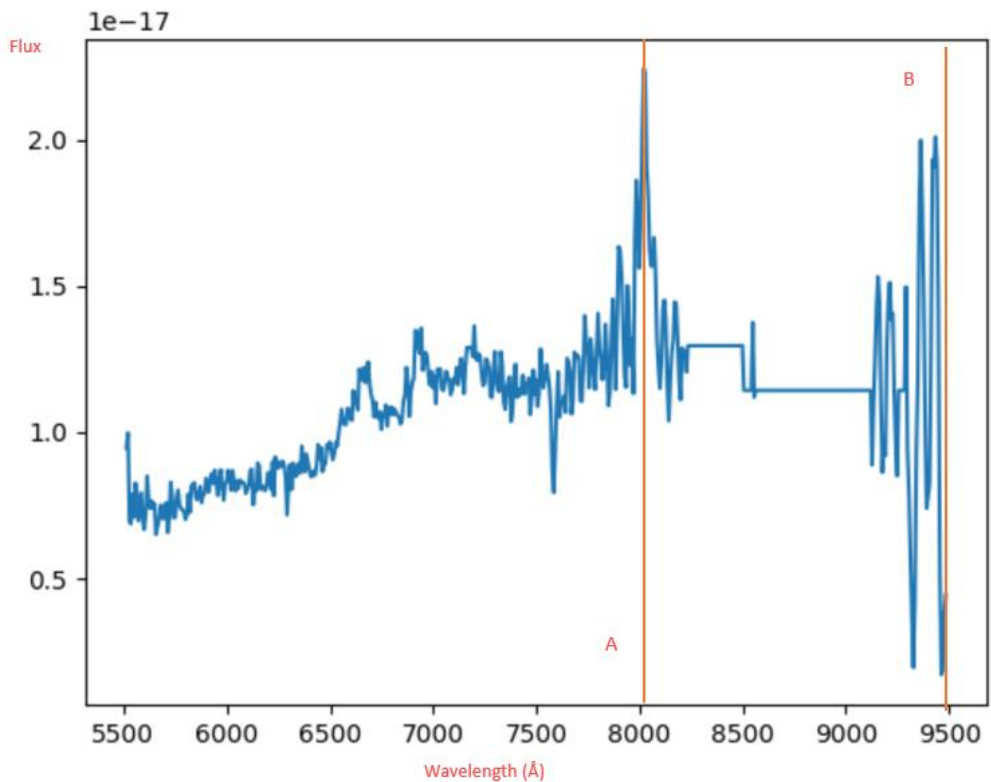

The most likely cause of blueshifting of emission lines is due to multi-body system ejections as AGNs were shown to eject quasars due to multi-body system interactions (Bell 2004). These ejections can occur at high velocities that causes a blueshifting of the emission lines if it is ejected towards us, the observer rather than away (Valtonen & Basu 1991). Blueshift also differs from redshift in its relationship with velocity. Due to the universal speed limit of the speed of light, there is a maximum value of blueshift of 1 since the doppler velocity cannot exceed the speed of light, whereas, redshift does not have an upper limit, it can increase continuously without the recessional velocity ever reaching the speed of light as seen in Figure 6.

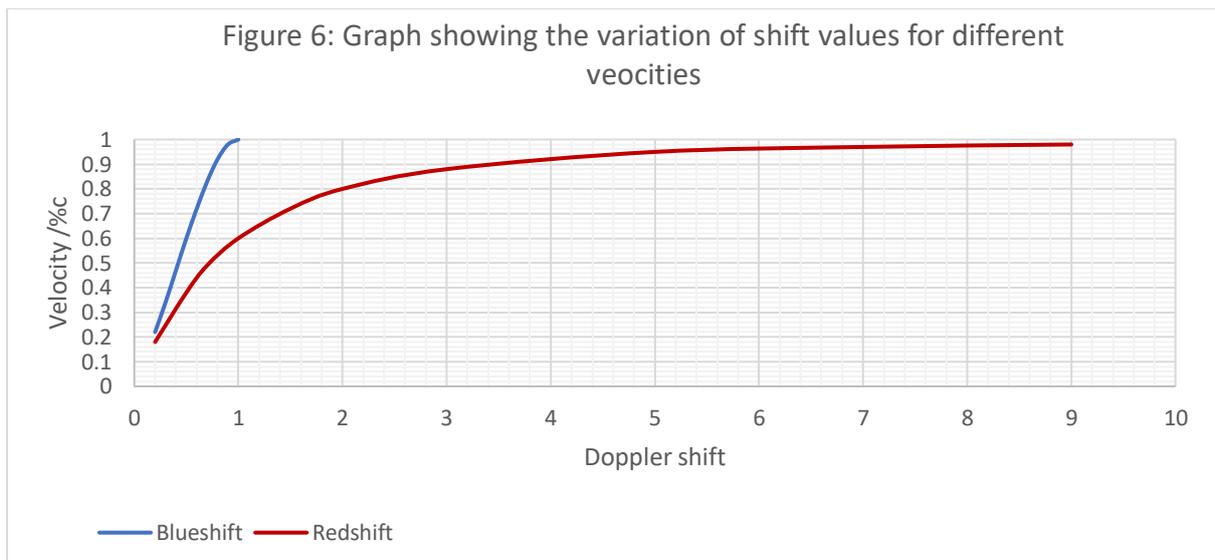

It should be noted, however, that there are some limitations of blueshift interpretation, as there is a maximum blueshift that can be detected based on the range of the spectra and the available search lines. The maximum possible blueshift in this case is 0.7624, based on having two strong emission lines present. This will give a maximum velocity of 0.89c (Valtonen & Basu 1991). This value falls within the range of previously found ejection velocities of up to 0.94c (Narlikar & Edmunds 1981). The largest blueshift value obtained in this analysis was 0.6241, this gives a maximum velocity of 0.75c. These maximum velocities are based on the blueshift values, while they are within the possible ejection velocities, to be moving towards us, the observer, at such a high velocity will actually require more than this measured velocities. The ejection velocity will have to be great enough to overcome the redshift present in the expansion of the universe and still have a large enough velocity to incur a blueshift. While these large blueshifts may not be entirely possible, depending on their distances away, 94% of our values have blueshifts smaller than 0.5. This also calls for a reconsideration of the distance measures for these quasars, if their distances were determined independent of their redshifts or not.

5) **Conclusion**

Quasars' spectra have been studied for a long time, but analysis of these spectra have always been under the redshift interpretation. While providing satisfactory answers for most spectra, the redshift interpretation is still plagued by inconsistencies for others. The blueshift interpretation can better explain some these spectra, without encountering the same inconsistencies as the redshift interpretation.

While blueshift is not as widespread as redshift in the universe, due to the general expansion of the universe, it is still a likely scenario. Even objects that are not as close to us, the observer, can experience blueshifts due to multi-body interactions. Chitan et al (2022) showed that a significant amount of SMBH triple systems result in ejections, and if these ejections are toward us, they can be travelling at velocities high enough to be blueshifted.

From the results gathered, blueshift can appropriately explain spectra that are not analyzed with lines of appropriate strengths using redshift. We have also shown a significant decrease in the number of unidentified emission lines of spectra under the blueshift interpretation compared to the redshift.

These are some of the reasons to give the blueshift hypothesis serious consideration for some quasar spectra and to help fill the gaps in redshift analysis of quasar spectra especially in light of recent discoveries.